\documentclass[twocolumn]{aastex63}
\usepackage[utf8]{inputenc}

\shorttitle{CEMP Stars in the Disk System of the Milky Way}
\shortauthors{Dietz et al.}

\begin{document}

\title{Two Populations of Carbon-Enhanced Metal-Poor Stars in the Disk System of the Milky Way}
\author[0000-0002-3567-1897]{Sarah E. Dietz}
\affiliation{Department of Physics and JINA Center for the Evolution of the Elements, University of Notre Dame, Notre Dame, IN 46556, USA}

\correspondingauthor{Sarah E. Dietz}
\email{sdietz@nd.edu}

\author[0000-0002-4168-239X]{Jinmi Yoon}
\affiliation{Department of Physics and JINA Center for the Evolution of the Elements, University of Notre Dame, Notre Dame, IN 46556, USA}

\author[0000-0003-4573-6233]{Timothy C. Beers}
\affiliation{Department of Physics and JINA Center for the Evolution of the Elements, University of Notre Dame, Notre Dame, IN 46556, USA}

\author[0000-0003-4479-1265]{Vinicius M. Placco}
\affiliation{Department of Physics and JINA Center for the Evolution of the Elements, University of Notre Dame, Notre Dame, IN 46556, USA}
\affiliation{NSF's National Optical-Infrared Astronomy Research Laboratory, Tucson, AZ 85719, USA}

\author[0000-0001-5297-4518]{Young Sun Lee}
\affiliation{Department of Astronomy and Space Science, Chungnam National University, Daejeon 34134, Republic of Korea}

%\author{The AEGIS Collaboration}

\begin{abstract}

We present a chemo-dynamical analysis of low-resolution ($R \sim 1300$) spectroscopy of stars from the AAOmega Evolution of Galactic Structure (AEGIS) survey, focusing on two key populations of carbon-enhanced metal-poor (CEMP) stars within the disk system of the Milky Way: a mildly prograde population ($L_z < 1000$\,kpc\,km\,s$^{-1}$) and a strongly prograde ($L_z > 1000$\,kpc\,km\,s$^{-1}$) population. Based on their chemical and kinematic characteristics, and on comparisons with similar populations found in the recent literature, we tentatively associate the former with an ex-situ inner-halo population originating from either the $Gaia$ Sausage or $Gaia$-Enceladus. The latter population is linked to the metal-weak thick-disk (MWTD). We discuss their implications in the context of the formation history of the Milky Way.\\

\end{abstract}

\submitjournal{\apj}

%\keywords{Galaxy: disk - Galaxy: evolution - Galaxy: formation stars: carbon - stars: kinematics and dynamics - methods: statistical}

\section{Introduction}\label{sec:intro}

The Milky Way's disk system is the most highly populated region of our Galaxy, and our position within this system enables the accumulation of a wealth of data to produce highly detailed characterizations to compare with numerical simulations of the thin- and thick-disk populations.

The thick-disk component was first formally proposed by \citet{yoshii_1982} and confirmed by \cite{gilmore_1983}, who demonstrated the need for an additional disk component when constructing Galactic stellar-density models. Since then, studies have uncovered rich substructure within the disk system, including the identification \citep{morrison_1990} and subsequent confirmation \citep{chiba_2000,beers_2014} of the metal-weak thick disk (MWTD). However, for almost three decades, despite numerous analyses, it remained unclear whether the MWTD was a separate population, or the metal-poor tail of the canonical thick disk.   This situation may now be resolved; two recent analyses indicate that the MWTD comprises a distinct component with its own unique formation history. \cite{carollo_2019} used a sample of 9,258 local stars from the Sloan Extension for Galactic Understanding and Exploration (SEGUE; \citealt{yanny_2009}) program of the Sloan Digital Sky Survey (SDSS; \citealt{york_2000}) to separate the MWTD from the thick-disk, finding the two populations to possess different characteristic kinematics, metallicities, and $\alpha$-element abundances. \cite{an_2020} constructed a chemo-dynamical ``blueprint" of Galactic components using photometric data from SDSS DR14 supplemented with deeper $u$-band photometry from the South Galactic Cap u-band Sky Survey (SCUSS; \citealt{gu_2015}) and astrometry from $Gaia$ DR2 \citep{gaia_dr2}, which is less subject to bias compared to targeted spectroscopic data. These authors identified several key stellar populations in their chemo-dynamical maps, including a MWTD component that is clearly separable from the canonical thick-disk stellar population.

The origin story for the disk system has also become more complex with the discovery of a relatively massive accreted satellite, known alternatively as the $Gaia$ Sausage or $Gaia$-Enceladus (the exact characteristics and potentially overlapping origins of these two proposed progenitors are still under debate; see, e.g., \citealt{evans_2020}), which may have contributed to the formation of the thick disk via dynamical heating as it merged with the Milky Way \citep{belokurov_2018,helmi_2018}.  The identification of a Splashed Disk population of stars \citep{belokurov_2020,an_2020} that may be connected with the proposed satellite collision(s) contributes an additional feature that could help constrain models for the formation of the disk system.

Recent reports of larger-than-expected populations of metal-poor stars within the disk system are also raising new questions about the assembly history of the Galaxy. The thin- and thick-disk metallicity distribution functions (MDFs) peak at approximately [Fe/H]\footnote{[A/B]$\equiv \log_{10}(N_{\rm A}/N_{\rm B})_* - \log_{10}(N_{\rm A}/N_{\rm B})_\odot$, where $N_{\rm A}$ and $N_{\rm B}$ are the number densities of elements A and B, respectively.} = $-0.1$ and [Fe/H] = $-0.6$, respectively, with the MWTD covering an approximate range of $-1.8 <$ [Fe/H] $< -0.8$ \citep{carollo_2007,carollo_2010}. However, \cite{sestito_2019} identified a significant population of ultra metal-poor stars (UMP; [Fe/H] $< -4.0$), well outside of the disk system's usual metallicity range, traveling on prograde orbits within 3\,kpc of the Galactic plane. They followed-up on this finding in \cite{sestito_2020}, using a combined sample of 1,027 very metal-poor (VMP) stars with [Fe/H] $<$ $-2.5$, observed with the Large Sky Area Multi-Object Fibre Spectroscopic Telescope (LAMOST; \citealt{LAMOST}) and the Pristine survey \citep{pristine_iii,pristine_vi}, demonstrating a statistically significant over-density of prograde VMP stars residing in the disk region. Similarly, \citealt{cordoni_2020} find $\sim$11\% of their 475 VMP stars from the SkyMapper survey \citep{skymapper_dr1} are within 3\,kpc of the plane and have prograde orbits with low eccentricities. \citealt{matteo_2020} even find an ``ultra metal-poor thick disk", extending as far down as [Fe/H] $\sim$ $-$6, within their sample of 54 VMP stars from the ESO Large Programme ``First Stars" \citep{ESO_LP_firststars}, with interesting implications for the early dynamical history of the Galaxy.

Complementary to these discoveries of metal-poor disk populations, numerous carbon-enhanced metal-poor (CEMP; [Fe/H] $< -1$, [C/Fe] $> +0.7$) stars have been identified in the disk system as well. In their analyses of metal-poor stars from the Hamburg/ESO survey, \cite{beers_2017} noted a population of CEMP-$s$\footnote{CEMP-$s$ stars exhibit over-abundances of elements associated with the slow neutron-capture process: [Ba/Fe] $>$ +1.0, [Ba/Eu] $>$ +0.5 \citep{beers_2005}.} stars in a kinematic and metallicity region usually associated with the MWTD. Yoon et al. (in prep.) find preliminary results indicating significant populations of CEMP stars in regions of energy-momentum space associated with the disk system, including a prograde population and a population with little to no angular momentum. Most notably, their sample includes a subset of ultra metal-poor (UMP; [Fe/H] $<$ $-4$) CEMP-no\footnote{CEMP-no stars exhibit no over-abundances of elements associated with neutron-capture processes: [Ba/Fe] $<$ 0 \citep{beers_2005}.} stars mainly found within the low-angular momentum population. These differences in kinematic and chemical characteristics suggest that at least two separate formation scenarios (e.g., from two accretion events) may be necessary to explain the presence of the CEMP stars in the disk populations.

In this paper, we continue this study of disk-like CEMP stars using low-resolution ($R \sim 1,300$) spectroscopy obtained by the AAOmega Evolution of Galactic Structure (AEGIS) survey (P.I. Keller), originally commissioned to study the evolutionary history of the thick-disk and halo systems of the Milky Way. As we demonstrate below, this sample includes two relatively nearby populations of CEMP stars, with potential implications for our understanding of the formation histories of the canonical thick disk and MWTD.
We introduce the AEGIS dataset in Section \ref{sec:data}, and describe its chemical abundances (Section \ref{subsec:chem}) and kinematics (Sections \ref{subsec:kin_dat} and \ref{subsec:kin_deriv}). Section \ref{sec:analysis} presents our analyses of this sample with results. We discuss the implication of our results in the context of the Galactic formation history in Section \ref{sec:results}. A brief summary of this work and our key findings are provided in Section \ref{sec:summ}.

\section{Data}\label{sec:data}

AEGIS is a spectroscopic survey conducted at the Australian Astronomical Telescope (AAT), using the dual beam (blue and red arms, covering ranges $\lambda=3,700$ to 5,800\,{\AA} and $\lambda=8,8400$ to 8,800\,{\AA}) AAOmega multi-object spectrograph to target populations of interest selected from the SkyMapper photometric survey. The resulting dataset comprises $\sim$70,000 stars with low-resolution spectroscopy ($R \sim 1,300$ for blue-arm spectra, $R \sim 10,000$ for red-arm spectra) and spans $\sim$4,900\,deg.$^2$ of sky in the Southern Hemisphere. A more complete description of the dataset can be found in \cite{yoon_2018}, along with a detailed examination of the metallicity ([Fe/H]) and carbonicity ([C/Fe]) of the Galactic halo through the lens of the AEGIS survey. 

\subsection{Chemical Abundances}\label{subsec:chem}

Stellar atmospheric parameters and a limited set of chemical abundances were derived with the non-SEGUE stellar parameter pipeline (n-SSPP; \citealt{beers_2014,beers_2017}). Effective temperature ($T_{\rm eff}$), surface gravity (log $g$), metallicity ([Fe/H]), and carbon abundances ([C/Fe]) for the AEGIS sample have been corrected to be more consistent with external high-resolution estimates, following the procedure described in \citet{beers_2014}. Additionally, we apply the evolutionary carbon corrections developed by \cite{placco_2014} to take into account the surface carbon-abundance depletion expected to occur on the upper red giant branch. For this sample, mean errors on $T_{\rm eff}$, log $g$, [Fe/H], and [C/Fe] are approximately 75\,K, 0.2\,dex, 0.1\,dex, and 0.1\,dex, respectively.

In the past, high-resolution spectroscopy was required to divide CEMP stars into the CEMP-no and CEMP-s sub-classes, because it was necessary to obtain barium and europium abundances to do so. However, \citet{yoon_2016} showed that this separation can be reliably (with a $\sim$90\% success rate) made using only the absolute carbon abundance, $A$(C)\footnote{$A$(C) $=\log \epsilon$(C) = log ($N_C/N_H$) + 12, where N indicates the number density of each species.}, making larger, low/medium-resolution datasets available for analysis. Because the division between CEMP-no and CEMP-$s$ stars can vary based on temperature and luminosity class, here we limit ourselves to the two categories for which the $A$(C) divisions are most apparent in this sample: 1) giants and sub-giants (G/SG) and 2) main-sequence dwarfs and turn-off stars (D/TO). We use divisions of $A$(C) $=7.1$ and $A$(C) $=7.6$ for the G/SG and D/TO classes, respectively, as suggested by \cite{yoon_2018} in their analysis of the AEGIS dataset. After removing duplicate measurements and measurements with signal-to-noise ratios $<$10, there are 1,061 G/SG CEMP stars and 421 D/TO CEMP stars identified in the AEGIS sample in total. The combined sample of these classes comprises 660 CEMP-no and 822 CEMP-$s$ stars.

We note that stellar temperature can affect our ability to adequately derive a star's carbon abundance. Compared to stars with strong carbon enhancements, those with moderate carbon enhancements can be difficult to detect in warmer ($T_{\rm eff} \gtrsim 5750$\,K) stars, producing a spurious over-abundance of high-$A$(C), high-$T_{\rm eff}$ stars (in other words, a higher CEMP-$s$ to CEMP-no ratio). The application of such a cut on temperature would substantially reduce our CEMP sample size, so we choose to present the sample without temperature restriction in the following analyses,  but make note of the effects that a temperature limit might have on our results, where appropriate.

\subsection{Kinematic Parameters}\label{subsec:kin_dat}

Radial velocities were derived using the n-SSPP analysis of the high-resolution red arm of the AEGIS spectra. A correction of $-24.6$\,km\,s$^{-1}$ was applied to all radial velocity values to account for an offset between the n-SSPP values and radial velocities derived using Ca triplet lines (at $\lambda$ $=$ 8498, 8542, 8662\,{\AA}) from the red-arm spectra (Navin, C. A., private communication). Proper motions from $Gaia$ DR2 \citep{gaia_dr2} are available for the majority ($\sim$98\%) of the sample. For the remaining $\sim$2\%, proper motions were averaged from a variety of catalogues (including Hipparcos, Tycho-1, and Tycho-2, as described in \citealt{beers_2014}).  We adopt a $+0.054$ correction to all $Gaia$ parallaxes as prescribed by \citet{schonrich_2019}, and derive distances from the inverted parallaxes for all stars with $<$20\% relative parallax uncertainty ($\sim$54\% of the sample). The remaining $\sim$46\% of the stars in our sample are assigned photometrically derived distances, following the procedure outlined in \citet{beers_2000}, as modified by \citet{beers_2012}.

\subsection{Kinematic Derivations}\label{subsec:kin_deriv}

Galactocentric positions and velocities are derived using the \texttt{galpy} Galactic dynamics package \citep{galpy}. In this work, we use $R_\odot$ = 8.2\,kpc for the distance to the center of the Galaxy \citep{bland_hawthorn_2016}, $v_{\rm LSR}$ = 236\,km\,s$^{-1}$ for the local standard of rest (LSR) velocity \citep{kawata_2019}, and ($U$, $V$, $W$)$_\odot$ = ($-$11.10,12.24,7.25)\,km\,s$^{-1}$ for the motion of the Sun with respect to the LSR \citep{schonrich_2010}. Orbital parameters are derived with the Galactic potential code used by \cite{chiba_2000}, which adopts the St\"ackel potential described in \cite{sommer-larsen_1990}.

To estimate uncertainties on the orbital parameters, we follow a Monte Carlo sampling procedure, as in \citet{dietz_2020}. A new set of kinematic input parameters is selected for each star from a corresponding set of random distributions (with the assumption that the given kinematic uncertainties are normally distributed about their observed values). This process is repeated 1,000 times per star, and the standard deviations of the resulting orbital-parameter distributions are taken as the uncertainties.

We note here that this sampling process should take into account the correlations between the input parameters in order to derive the most accurate uncertainty. However, correlation coefficients are not available for the kinematic parameters given in the original AEGIS dataset (as noted above, we use the original kinematic parameters given in the AEGIS dataset for 100\% of our radial velocities, $\sim$2\% of our proper motions, and $\sim$46\% of our distances). Including correlation coefficients in our calculations for stars with $Gaia$ kinematics results in (at most) a median difference of $\sim$1\% and mean difference of $\sim$6\% in derived uncertainties for the orbital parameters used in this work when compared to uncertainties calculated without correlation coefficients. Because this difference is minor, we choose to neglect correlations between input parameters in order to treat the sub-sets of our data with AEGIS and $Gaia$ kinematics in the same manner.

To avoid identifying any potentially spurious features, we limit our sample to stars with uncertainties on $Z_{\rm max}$ less than 1\,kpc and uncertainties on $L_z$ less than 250\,kpc\,km\,s$^{-1}$ (that is, no greater than our chosen bin size in Figure \ref{fig:hist_slices}). After applying this restriction, we have a total of 51,946 stars in the $Z_{\rm max} < 5$\,kpc region, 427 of which are CEMP-$s$ stars and 223 of which are CEMP-no stars.

\section{Analysis}\label{sec:analysis}

\begin{figure*}
\plotone{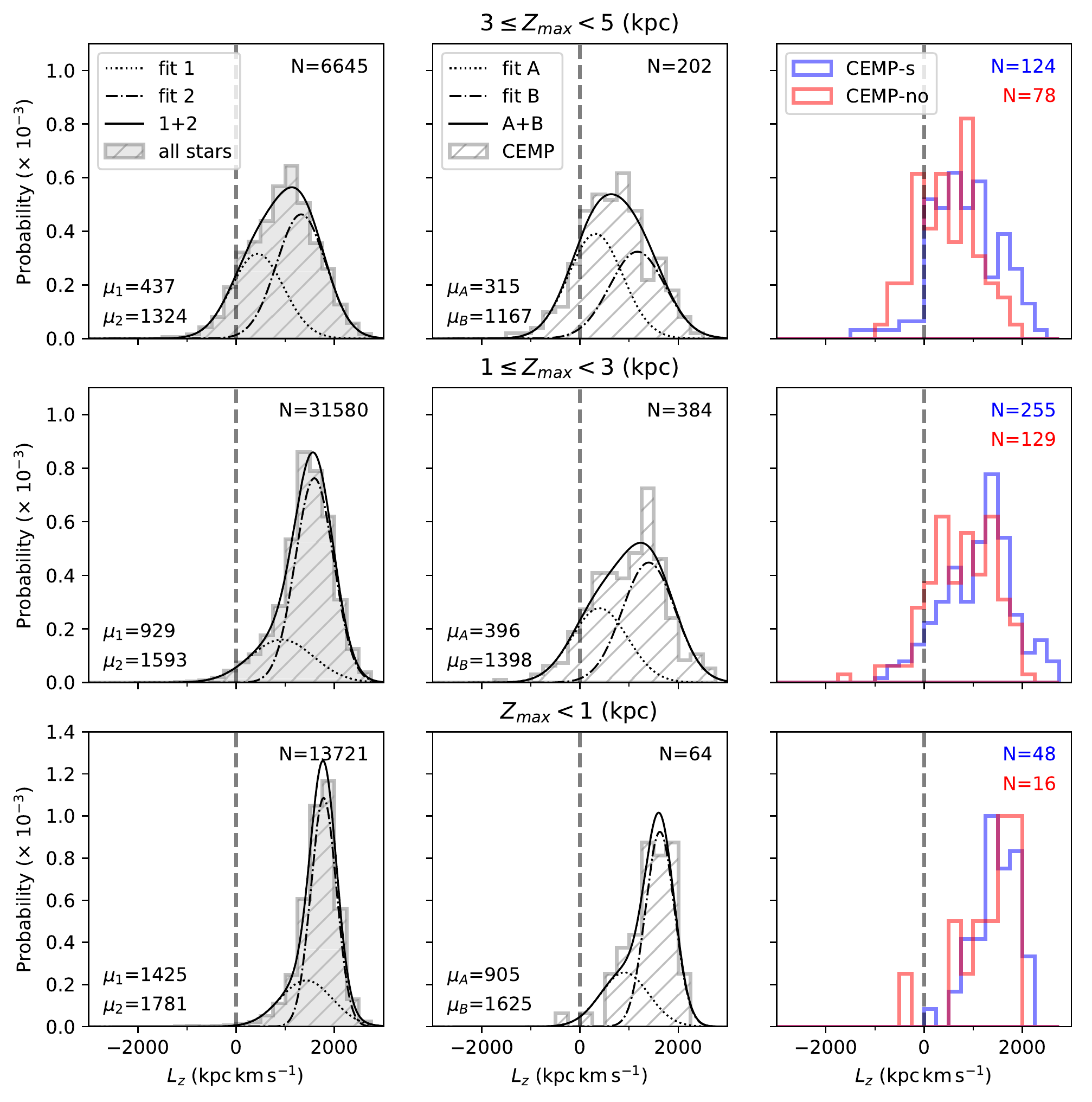}
\caption{Angular momentum distributions for the sample over three different ranges of $Z_{\rm max}$. The left column of panels shows all stars with valid kinematics, the middle column shows the subset of CEMP stars, and the right column shows the CEMP subset divided into CEMP-$s$ (blue) and CEMP-no (red) distributions. The total population (left) and the CEMP subset (middle) are each fit with two-component Gaussian distributions. The means of these fits are indicated in the bottom left-hand corner of these panels. The number of stars plotted, N, is given in the upper right-hand corner of each panel. For the right column, N is given for the CEMP-$s$ and CEMP-no subsets in blue and red, respectively. A dashed line marks $L_z = 0$\,kpc\,km\,s$^{-1}$ for reference in each plot.}
\label{fig:hist_slices}
\end{figure*}

We begin our analysis by identifying populations of interest close to the Galactic plane. Angular momentum ($L_z$) distributions for the sample are divided into sections based on maximum orbital extent from the Galactic plane, $Z_{\rm max}$, as shown in Figure \ref{fig:hist_slices}.

From left to right, the columns of Figure \ref{fig:hist_slices} show the distributions for all stars, the CEMP stars, and the CEMP-$s$ (blue) +  CEMP-no (red) stars.

In the full sample (left column of panels), the disk clearly dominates at all $Z_{\rm max}$ ranges, producing a strongly prograde peak at $L_z>1000$\,kpc\,km\,s$^{-1}$. This peak includes both thin- and thick-disk stars, but it should be noted that the thin disk is not fully represented here due to the metallicity upper limit within the AEGIS sample ([Fe/H] $\leq$ 0.3). The inner-halo component ($L_z\sim0$\,kpc\,km\,s$^{-1}$) becomes more visible at $3\leq Z_{\rm max}<5$\,kpc, although the disk system still retains a robust peak even at these heights.

In the CEMP sub-sample (middle column of panels in Figure \ref{fig:hist_slices}), at least two populations appear to be present for all $Z_{\rm max}$ ranges. We have fitted the $L_z$ distributions with Gaussians using the \texttt{scikit-learn} mixture package in order to approximate the general features of these populations (we have also performed similar fits on the total sample so that we can compare the characteristics of the total sample to the CEMP sub-samples). Each range contains a mildly prograde peak and a strongly prograde peak---we  refer to these as populations ``A" and ``B", respectively, for the remainder of this work. The low-momentum peak is likely associated with the inner-halo population, a rich source of CEMP stars, which would account for the larger relative proportion of population A at high $Z_{\rm max}$. Population B displays a strong net rotation and decreases in relative significance with increasing $Z_{\rm max}$, which suggests it may be a part of thick-disk/MWTD.

The fits for population A peak at 905, 396, and 315\,kpc\,km\,s$^{-1}$, from the low to high $Z_{\rm max}$ ranges. The fits for population B peak at 1625, 1398, and 1167\,kpc\,km\,s$^{-1}$, from the low to high $Z_{\rm max}$ ranges. These fits are mainly meant to provide an overview of the characteristics of our CEMP populations, not to create a strict definition for each population, so it is understandable that the location of the peaks varies somewhat with $Z_{\rm max}$ (especially at $Z_{\rm max} < 1$\,kpc, where population A is weakly represented). It is interesting to note here that population B lags an average of $\sim 170$\,kpc\,km\,s$^{-1}$ behind the dominant, strongly prograde peak of the total sample.

Populations A and B appear to possess different relative fractions of CEMP-$s$ and CEMP-no stars, as can be seen in the right column of panels in Figure \ref{fig:hist_slices}. The CEMP-$s$ to CEMP-no ratios for each population are listed in Table \ref{tab:hist_ratios}, as well as a complementary set of ratios for a limited-temperature cut of the sample (see Section \ref{subsec:chem} for details on the effect of $T_{\rm eff}$ on relative CEMP ratios). These ratios are obtained using \texttt{scikit-learn} mixture package, which allows us to predict  population membership fractions for the CEMP stars in our sample based on our fits by predicting which Gaussian component each input star most probably belongs to.

Both populations are dominated by CEMP-$s$ stars, which is not surprising, given that we currently understand CEMP-no stars to have predominantly ex-situ origins \citep[e.g.,][]{lee_2017,lee_2019,yoon_2018, yoon_2019,yoon_2020}, though the relative strength of this ratio appears to vary based on the sub-sample being considered. In the full sample, the ratio of CEMP-$s$ to CEMP-no stars is roughly twice as large in population B as it is in population A, which could suggest different origins for the CEMP stars within these populations.

When we consider the sample restricted to $T_{\rm eff}<5750$\,K, CEMP-$s$ stars still dominate both populations, but the CEMP-$s$ to CEMP-no ratio varies much more unpredictably, making it challenging to make any definitive statement on the chemical origins of population A versus population B. Note that the low-temperature sample contains significantly fewer CEMP stars than the full sample; a larger sample of cool CEMP stars in this region may be needed to more fully explore these populations.

\begin{deluxetable*}{lclll}
\tablecaption{CEMP-$s$ to CEMP-no ratios (and total CEMP counts) for Figure \ref{fig:hist_slices} \label{tab:hist_ratios}}
\tablehead{
\colhead{} & Pop. & \colhead{$0<Z_{\rm max}\leq 1$\,kpc} & \colhead{$1<Z_{\rm max}\leq 3$\,kpc} & \colhead{$3<Z_{\rm max}\leq 5$\,kpc}
}
\startdata
All $T_{\rm eff}$ & A & \textcolor{blue}{1.8} (17) & \textcolor{blue}{1.3} (143) & \textcolor{blue}{1.1} (110) \\
{} & B & \textcolor{blue}{3.7} (47) & \textcolor{blue}{2.7} (241) & \textcolor{blue}{2.5} (92) \\
\hline
$T_{\rm eff}<5750$\,K & A & \textcolor{blue}{3.7} (14) & \textcolor{blue}{1.9} (60) & \textcolor{blue}{1.1} (68) \\
{} & B & \textcolor{blue}{2.0} (6) & \textcolor{blue}{2.1} (56) & \textcolor{blue}{1.8} (14) \\
\enddata
\vspace{0.15in}
\tablecomments{The CEMP-$s$ to CEMP-no ratios for populations A and B are given in blue for each range shown in Figure \ref{fig:hist_slices}, for the full sample and for a temperature-limited sample. The total number of CEMP stars in each population for the given range is listed in parentheses.}
\end{deluxetable*}

\begin{figure*}
\plotone{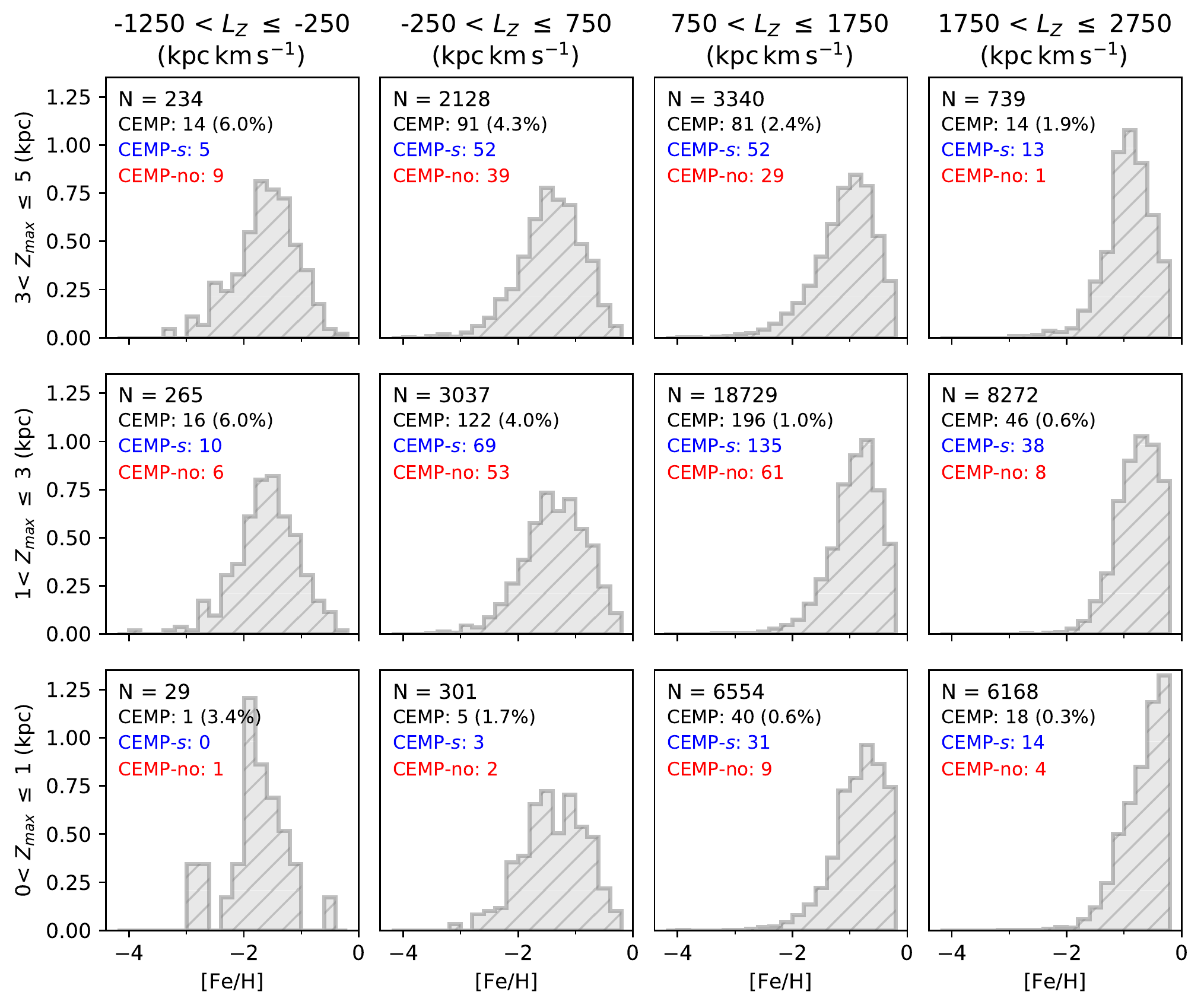}
\caption{Normalized MDFs for the sample over three different ranges of $Z_{\rm max}$ (rows) and four different ranges of $L_Z$ (columns). The total star count (N) is noted in the upper left-hand corner of each plot. The total CEMP count and the percentage of CEMP stars (relative to total count) is listed below N. CEMP-$s$ and CEMP-no counts are given in blue and red, respectively.}
\label{fig:MDFs}
\end{figure*}

The presence of a large number of CEMP stars in a region of the Galaxy usually associated with disk stars is worthy of further investigation. To aid in interpretation of these data, we present the same samples of stars shown in Figure \ref{fig:hist_slices} in a set of MDFs in Figure \ref{fig:MDFs}. Rows are sub-divided into the same $Z_{\rm max}$ ranges used in Figure \ref{fig:hist_slices}, while columns are separated into $L_Z$ ranges. Population counts and statistics are given in the upper left-hand corner of each sub-plot. 

Inspection of Figure \ref{fig:MDFs} shows that the strongly prograde stars in our sample are generally more metal rich than the mildly prograde or retrograde stars, as expected for a disk-dominated sample. As in Figure \ref{fig:hist_slices}, the disk is robustly represented (high metallicity, strongly prograde) at both low and high $Z_{\rm max}$, and here too we observe the growing inner-halo contribution ([Fe/H] $\sim -1.6$, $L_z \sim 0$\,kpc\,km\,s$^{-1}$) in the highest $Z_{\rm max}$ range.

Figure \ref{fig:MDFs} also includes CEMP, CEMP-$s$, and CEMP-no counts for the each kinematic range, listed in black, blue, and red, respectively. The relative percentage of CEMP stars compared to all stars is noted in parentheses next to the CEMP count. Although the strongly prograde stars (two right-most columns) have the most CEMP stars by number, they possess the smallest relative percentages of CEMP stars compared to the total population. We find a relatively large number of CEMP stars in these regions simply because these regions of the kinematic space were sampled the most in the observations. Nevertheless, the presence of even a small relative percentage of CEMP stars moving in tandem with the disk is interesting, and may provide insight into the disk's formation history. These sub-samples correspond to the CEMP-$s$-rich population B noted above.

Population A can be seen more clearly in the lower-$L_Z$ ranges (two left-most columns). These regions contain small absolute numbers of CEMP stars, but possess the highest CEMP percentages. A feature of note here is the double metallicity peak seen in both the $Z_{\rm max}\leq 1$\,kpc and $1 < Z_{\rm max}\leq 3$\,kpc plots within the $-250 < L_Z \leq 750$\,kpc\,km\,s$^{-1}$ range. Peaks at approximately [Fe/H] $= -1.0$ and [Fe/H] $= -1.7$ are present in the $Z_{\rm max}\leq 1$\, kpc sub-sample, becoming less distinct as we move farther from the plane. It should be noted that the shape of this component varies somewhat with binning, though a feature similar to the [Fe/H] $= -1.0$ peak can also be seen in Figure 4 of \citealt{an_2020}; the authors suggest the Splashed Disk, presented in \cite{belokurov_2020}, as one possible source.

\section{Discussion}\label{sec:results}

We have identified two CEMP populations of interest in the disk system of the Milky Way: the mildly prograde population A ($L_z < 1000$\,kpc\,km\,s$^{-1}$) and the strongly prograde population B ($L_z > 1000$\,kpc\,km\,s$^{-1}$), both containing an enhancement of CEMP-$s$ stars relative to CEMP-no stars.

Although many population A stars orbit close to the Galactic plane, this population may be linked to the inner-halo population, particularly since it possesses a similar relative percentage of CEMP-$s$ stars (53-65\%, depending on $Z_{\max}$) to that given by \cite{carollo_2014} for this component (57\%). \cite{an_2020} found a strong inner-halo population even at slices of $|Z|$ close to the plane, estimating two-thirds of the metal-poor stars in the $1<|Z|<2$\,kpc region of their data to be $Gaia$-Enceladus stars. Although the mildly prograde motion of population A is at odds with the slightly retrograde motion derived by \citet{helmi_2018} for $Gaia$-Enceladus, a common origin cannot be ruled out. Both population A and $Gaia$-Enceladus span a range of velocities, including both prograde and retrograde rotation, and the latter presumably carries a similar CEMP-$s$ percentage to that quoted in \citet{carollo_2014}, as $Gaia$-Enceladus is proposed to make up a large portion of the inner-halo population. It is also possible that population A is instead a part of the $Gaia$ Sausage, which possesses a slightly higher mean $L_z$ ($L_z \sim 0$\,kpc\,km\,s$^{-1}$) than $Gaia$-Enceladus (bounded by $-1,500$\,kpc\,km\,s$^{-1}$ $< L_z < 500$\,kpc\,km\,s$^{-1}$ in \citealt{helmi_2018}). The scientific community has not yet come to a consensus on which scenario better describes the formation of the ex-situ inner-halo population, the $Gaia$ Sausage or $Gaia$-Enceladus; it would be interesting to revisit the characteristics of population A in the future, when more is known about the nature of the main inner-halo progenitor(s).

Population B possesses kinematic characteristics more in-line with the thick-disk system ($L_z \sim 1,500$\,kpc\,km\,s$^{-1}$, close to the Galactic plane), and the low metallicity of our CEMP stars (by definition) necessarily designate them as members of the MWTD, which spans an approximate range of $-1.8 <$ [Fe/H] $< -0.8$ \citep{carollo_2010}. \cite{beers_2017} and Yoon et al. (in prep.) also noted significant CEMP-$s$ populations in MWTD-associated regions of their samples. It is unclear whether these stars formed in-situ or were imported into the disk system. CEMP stars are not expected to be common in a well-mixed, gas-rich environment like the disk, but peak B makes up a very small percentage of the total disk-system stars within its kinematic region, so in-situ formation is not out of the question. For instance, \cite{sestito_2020} propose a possible in-situ formation pathway for their population of disk VMP stars, involving pockets of pristine gas in the proto-disk and radial migration. On the other hand, both \cite{carollo_2019} and \cite{an_2020} find evidence in their data clearly indicating a separate MWTD population, which suggests a potential ex-situ origin for population B stars. \citet{lian_2020} propose a two-pronged formation scenario for the thick disk, including a late starburst in the outer disk, potentially caused by the accretion of a gas-rich dwarf galaxy. Although the abundance-space explored in their analyses ([Fe/H] $>$ $-1$) does not extend to the low-metallicity regimes probed here, it is possible that this ex-situ outer thick disk is linked to population B. In the case of an accreted origin, the high relative fraction of CEMP-$s$ stars in population B could indicate a (relatively) massive, gas-rich progenitor satellite, which would have preferentially formed more CEMP-$s$ stars than CEMP-no stars, the latter being mostly accreted from less-massive progenitors such as UFDs \citep{yoon_2019}.

An investigation into the morphological groups introduced by \citet{yoon_2016} present in our CEMP sub-populations could be of interest, especially an analysis of the two classes dominated by CEMP-no stars, ``Group II" and ``Group III". These groups are thought to have different progenitors due to their distinct $A$(C)-[Fe/H] and $A$(C)-$A$(Na, Mg) relations, which could provide insight into the origins of population A versus B, but our sample does not possess sufficient numbers of potential Group III stars (which can be difficult to identify, due the overlap between Groups II and III) to make any statistically interesting statements about the Group II/Group III ratio in either population. However, Yoon et al. (in prep.) find a strong Group III population in a region of energy-momentum space potentially associated with population A (low-energy, $L_z < 1000$\,kpc\,km\,s$^{-1}$) based on a high-resolution literature sample of Group III CEMP-no stars. \cite{yoon_2019} found Group III stars to be preferentially accreted from UFDs, so further sampling of the population A region may help constrain the assembly history of the nearby halo, as well as potentially contribute to the as-yet sparsely populated Group III region of the $A$(C)-[Fe/H] space.

\section{Summary and Conclusions}\label{sec:summ}

We present a chemo-dynamical analysis of $Z_{max} < 5$\,kpc stars from the AEGIS survey, focusing on CEMP populations within this region. We find two key CEMP populations of interest close to the Galactic plane: a mildly prograde ($L_z < 1000$\,kpc\,km\,s$^{-1}$) population (population ``A") and a strongly prograde ($L_z > 1000$\,kpc\,km\,s$^{-1}$) population (population ``B"). Population A contains a mild over-abundance of CEMP-$s$ compared to CEMP-no stars ($\sim$53-65\% CEMP-$s$), which, in combination with its kinematic characteristics (low $L_z$, dominant farther from the Galactic plane), lead us to associate this population with the inner-halo component. These stars could belong to either of the proposed ex-situ inner-halo progenitors: the $Gaia$ Sausage or $Gaia$-Enceladus. 

Population B also contains preferentially more CEMP-$s$ stars than CEMP-no stars (potentially with a higher ratio than population A), but a larger number of low-$T_{\rm eff}$, $Z_{\rm max} < 5$\,kpc CEMP stars than our current sample ($\sim$200) is needed to more fully explore this possibility), and can be kinematically and chemically associated with the MWTD. This clump of (mainly) CEMP-$s$ stars within the MWTD has been seen in other samples as well, including in \citet{beers_2017} and Yoon et al. (in prep.). We propose both in-situ and ex-situ origins for this population, such as pockets of pristine gas in the proto-disk (in-situ), as suggested by \citet{sestito_2020}, and a relatively massive merger of a gas-rich progenitor satellite (ex-situ).

Although the stellar halo (and the outer-halo component in particular) contains the highest relative ratio of metal-poor and CEMP stars compared to other Galactic components, a surprising number of these ancient tracer populations are emerging in recent surveys of the disk system. We present our own findings within the AEGIS dataset as potentially useful constraints for evolutionary models of the Milky Way, particularly with regards to the creation of the ex-situ inner-halo and the formation of the MWTD. Future surveys of the disk and halo systems will undoubtedly aid in interpretation of the CEMP behaviors noted here, and ongoing efforts to increase the number of known Group III stars could provide further constraints on the origins of these populations.

\acknowledgments

We thank all researchers involved in the AEGIS Collaboration for their efforts in observing and analyzing the data presented here, in particular Sarah Martell and Gary Da Costa for their feedback on this paper, as well as Colin Navin for deriving the radial velocity offsets used in this analysis.

This work has made use of data from the European Space Agency (ESA) mission
{\it Gaia} (\url{https://www.cosmos.esa.int/gaia}), processed by the {\it Gaia}
Data Processing and Analysis Consortium (DPAC,
\url{https://www.cosmos.esa.int/web/gaia/dpac/consortium}). Funding for the DPAC
has been provided by national institutions, in particular the institutions
participating in the {\it Gaia} Multilateral Agreement.

The authors acknowledge partial support
from grant PHY 14-30152, Physics Frontier Center/JINA Center for the
Evolution of the Elements (JINA-CEE), awarded by the US National Science
Foundation.

Y.S.L. acknowledges support from the National Research Foundation (NRF) of Korea grant funded by the Ministry of Science and ICT (No.2017R1A5A1070354 and NRF2018R1A2B6003961).

\software{galpy \citep{galpy}, numpy \citep{numpy}, matplotlib \citep{matplotlib}, scipy \citep{scipy}}

\bibliography{main.bib}

\begin{thebibliography}{}
\expandafter\ifx\csname natexlab\endcsname\relax\def\natexlab#1{#1}\fi
\providecommand{\url}[1]{\href{#1}{#1}}
\providecommand{\dodoi}[1]{doi:~\href{http://doi.org/#1}{\nolinkurl{#1}}}
\providecommand{\doeprint}[1]{\href{http://ascl.net/#1}{\nolinkurl{http://ascl.net/#1}}}
\providecommand{\doarXiv}[1]{\href{https://arxiv.org/abs/#1}{\nolinkurl{https://arxiv.org/abs/#1}}}

\bibitem[{{Aguado} {et~al.}(2019){Aguado}, {Youakim}, {Gonz{\'a}lez
  Hern{\'a}ndez}, {Allende Prieto}, {Starkenburg}, {Martin}, {Bonifacio},
  {Arentsen}, {Caffau}, {Peralta de Arriba}, {Sestito}, {Garcia-Dias},
  {Fantin}, {Hill}, {Jablonca}, {Jahandar}, {Kielty}, {Longeard}, {Lucchesi},
  {S{\'a}nchez-Janssen}, {Osorio}, {Palicio}, {Tolstoy}, {Wilson},
  {C{\^o}t{\'e}}, {Kordopatis}, {Lardo}, {Navarro}, {Thomas}, \&
  {Venn}}]{pristine_vi}
{Aguado}, D.~S., {Youakim}, K., {Gonz{\'a}lez Hern{\'a}ndez}, J.~I., {et~al.}
  2019, \mnras, 490, 2241, \dodoi{10.1093/mnras/stz2643}

\bibitem[{{An} \& {Beers}(2020)}]{an_2020}
{An}, D., \& {Beers}, T.~C. 2020, in American Astronomical Society Meeting
  Abstracts, American Astronomical Society Meeting Abstracts, 158.01

\bibitem[{{Beers} {et~al.}(2000){Beers}, {Chiba}, {Yoshii}, {Platais},
  {Hanson}, {Fuchs}, \& {Rossi}}]{beers_2000}
{Beers}, T.~C., {Chiba}, M., {Yoshii}, Y., {et~al.} 2000, \aj, 119, 2866,
  \dodoi{10.1086/301410}

\bibitem[{{Beers} \& {Christlieb}(2005)}]{beers_2005}
{Beers}, T.~C., \& {Christlieb}, N. 2005, \araa, 43, 531,
  \dodoi{10.1146/annurev.astro.42.053102.134057}

\bibitem[{{Beers} {et~al.}(2014){Beers}, {Norris}, {Placco}, {Lee}, {Rossi},
  {Carollo}, \& {Masseron}}]{beers_2014}
{Beers}, T.~C., {Norris}, J.~E., {Placco}, V.~M., {et~al.} 2014, \apj, 794, 58,
  \dodoi{10.1088/0004-637X/794/1/58}

\bibitem[{{Beers} {et~al.}(2012){Beers}, {Carollo}, {Ivezi{\'c}}, {An},
  {Chiba}, {Norris}, {Freeman}, {Lee}, {Munn}, \& {Re Fiorentin}}]{beers_2012}
{Beers}, T.~C., {Carollo}, D., {Ivezi{\'c}}, {\v{Z}}., {et~al.} 2012, \apj,
  746, 34, \dodoi{10.1088/0004-637X/746/1/34}

\bibitem[{{Beers} {et~al.}(2017){Beers}, {Placco}, {Carollo}, {Rossi}, {Lee},
  {Frebel}, {Norris}, {Dietz}, \& {Masseron}}]{beers_2017}
{Beers}, T.~C., {Placco}, V.~M., {Carollo}, D., {et~al.} 2017, \apj, 835, 81,
  \dodoi{10.3847/1538-4357/835/1/81}

\bibitem[{{Belokurov} {et~al.}(2018){Belokurov}, {Erkal}, {Evans}, {Koposov},
  \& {Deason}}]{belokurov_2018}
{Belokurov}, V., {Erkal}, D., {Evans}, N.~W., {Koposov}, S.~E., \& {Deason},
  A.~J. 2018, \mnras, 478, 611, \dodoi{10.1093/mnras/sty982}

\bibitem[{{Belokurov} {et~al.}(2020){Belokurov}, {Sanders}, {Fattahi}, {Smith},
  {Deason}, {Evans}, \& {Grand }}]{belokurov_2020}
{Belokurov}, V., {Sanders}, J.~L., {Fattahi}, A., {et~al.} 2020, \mnras,
  \dodoi{10.1093/mnras/staa876}

\bibitem[{{Bland-Hawthorn} \& {Gerhard}(2016)}]{bland_hawthorn_2016}
{Bland-Hawthorn}, J., \& {Gerhard}, O. 2016, \araa, 54, 529,
  \dodoi{10.1146/annurev-astro-081915-023441}

\bibitem[{Bonifacio {et~al.}(2009)Bonifacio, Andersen, Andrievsky, Barbuy,
  Beers, Caffau, Cayrel, Depagne, François, González Hernández, \&
  et~al.}]{ESO_LP_firststars}
Bonifacio, P., Andersen, J., Andrievsky, S.~M., {et~al.} 2009, Science with the
  VLT in the ELT Era, 31–35, \dodoi{10.1007/978-1-4020-9190-2_6}

\bibitem[{{Bovy}(2015)}]{galpy}
{Bovy}, J. 2015, \apjs, 216, 29, \dodoi{10.1088/0067-0049/216/2/29}

\bibitem[{{Carollo} {et~al.}(2014){Carollo}, {Freeman}, {Beers}, {Placco},
  {Tumlinson}, \& {Martell}}]{carollo_2014}
{Carollo}, D., {Freeman}, K., {Beers}, T.~C., {et~al.} 2014, \apj, 788, 180,
  \dodoi{10.1088/0004-637X/788/2/180}

\bibitem[{{Carollo} {et~al.}(2007){Carollo}, {Beers}, {Lee}, {Chiba}, {Norris},
  {Wilhelm}, {Sivarani}, {Marsteller}, {Munn}, {Bailer-Jones}, {Fiorentin}, \&
  {York}}]{carollo_2007}
{Carollo}, D., {Beers}, T.~C., {Lee}, Y.~S., {et~al.} 2007, \nat, 450, 1020,
  \dodoi{10.1038/nature06460}

\bibitem[{{Carollo} {et~al.}(2010){Carollo}, {Beers}, {Chiba}, {Norris},
  {Freeman}, {Lee}, {Ivezi{\'c}}, {Rockosi}, \& {Yanny}}]{carollo_2010}
{Carollo}, D., {Beers}, T.~C., {Chiba}, M., {et~al.} 2010, \apj, 712, 692,
  \dodoi{10.1088/0004-637X/712/1/692}

\bibitem[{{Carollo} {et~al.}(2019){Carollo}, {Chiba}, {Ishigaki}, {Freeman},
  {Beers}, {Lee}, {Tissera}, {Battistini}, \& {Primas}}]{carollo_2019}
{Carollo}, D., {Chiba}, M., {Ishigaki}, M., {et~al.} 2019, arXiv e-prints,
  arXiv:1904.04881.
\newblock \doarXiv{1904.04881}

\bibitem[{{Chiba} \& {Beers}(2000)}]{chiba_2000}
{Chiba}, M., \& {Beers}, T.~C. 2000, \aj, 119, 2843, \dodoi{10.1086/301409}

\bibitem[{{Cordoni} {et~al.}(2020){Cordoni}, {Da Costa}, {Yong}, {Mackey},
  {Marino}, {Monty}, {Nordlander}, {Norris}, {Asplund}, {Bessell}, {Casey},
  {Frebel}, {Lind}, {Murphy}, {Schmidt}, {Gao}, {Xylakis-Dornbusch}, {Amarsi},
  \& {Milone}}]{cordoni_2020}
{Cordoni}, G., {Da Costa}, G.~S., {Yong}, D., {et~al.} 2020, \mnras,
  \dodoi{10.1093/mnras/staa3417}

\bibitem[{{Cui} {et~al.}(2012){Cui}, {Zhao}, {Chu}, {Li}, {Li}, {Zhang}, {Su},
  {Yao}, {Wang}, {Xing}, {Li}, {Zhu}, {Wang}, {Gu}, {Luo}, {Xu}, {Zhang},
  {Liu}, {Zhang}, {Yang}, {Cao}, {Chen}, {Chen}, {Chen}, {Chen}, {Chu}, {Feng},
  {Gong}, {Hou}, {Hu}, {Hu}, {Hu}, {Jia}, {Jiang}, {Jiang}, {Jiang}, {Jin},
  {Li}, {Li}, {Li}, {Liu}, {Liu}, {Lu}, {Mao}, {Men}, {Qi}, {Qi}, {Shi},
  {Tang}, {Tao}, {Wang}, {Wang}, {Wang}, {Wang}, {Wang}, {Wang}, {Wang},
  {Wang}, {Wang}, {Wang}, {Wang}, {Wang}, {Xu}, {Xu}, {Yang}, {Yu}, {Yuan},
  {Yuan}, {Zhai}, {Zhang}, {Zhang}, {Zhang}, {Zhao}, {Zhou}, {Zhou}, {Zhu}, \&
  {Zou}}]{LAMOST}
{Cui}, X.-Q., {Zhao}, Y.-H., {Chu}, Y.-Q., {et~al.} 2012, Research in Astronomy
  and Astrophysics, 12, 1197, \dodoi{10.1088/1674-4527/12/9/003}

\bibitem[{{Di Matteo} {et~al.}(2020){Di Matteo}, {Spite}, {Haywood},
  {Bonifacio}, {G{\'o}mez}, {Spite}, \& {Caffau}}]{matteo_2020}
{Di Matteo}, P., {Spite}, M., {Haywood}, M., {et~al.} 2020, \aap, 636, A115,
  \dodoi{10.1051/0004-6361/201937016}

\bibitem[{{Dietz} {et~al.}(2020){Dietz}, {Yoon}, {Beers}, \&
  {Placco}}]{dietz_2020}
{Dietz}, S.~E., {Yoon}, J., {Beers}, T.~C., \& {Placco}, V.~M. 2020, \apj, 894,
  34, \dodoi{10.3847/1538-4357/ab7fa4}

\bibitem[{{Evans}(2020)}]{evans_2020}
{Evans}, N.~W. 2020, arXiv e-prints, arXiv:2002.05740.
\newblock \doarXiv{2002.05740}

\bibitem[{{Gaia Collaboration} {et~al.}(2018){Gaia Collaboration}, {Brown},
  {Vallenari}, {Prusti}, {de Bruijne}, {Babusiaux}, {Bailer-Jones}, {Biermann},
  {Evans}, {Eyer}, \& et~al.}]{gaia_dr2}
{Gaia Collaboration}, {Brown}, A.~G.~A., {Vallenari}, A., {et~al.} 2018, \aap,
  616, A1, \dodoi{10.1051/0004-6361/201833051}

\bibitem[{{Gilmore} \& {Reid}(1983)}]{gilmore_1983}
{Gilmore}, G., \& {Reid}, N. 1983, \mnras, 202, 1025,
  \dodoi{10.1093/mnras/202.4.1025}

\bibitem[{{Gu} {et~al.}(2015){Gu}, {Du}, {Jia}, {Peng}, {Wu}, {Jing}, {Ma},
  {Zhou}, {Fan}, {Fan}, {Jing}, {Jiang}, {Lesser}, {Nie}, {Shen}, {Wang},
  {Zou}, {Zhang}, \& {Zhou}}]{gu_2015}
{Gu}, J., {Du}, C., {Jia}, Y., {et~al.} 2015, \mnras, 452, 3092,
  \dodoi{10.1093/mnras/stv1529}

\bibitem[{{Helmi} {et~al.}(2018){Helmi}, {Babusiaux}, {Koppelman}, {Massari},
  {Veljanoski}, \& {Brown}}]{helmi_2018}
{Helmi}, A., {Babusiaux}, C., {Koppelman}, H.~H., {et~al.} 2018, \nat, 563, 85,
  \dodoi{10.1038/s41586-018-0625-x}

\bibitem[{Hunter(2007)}]{matplotlib}
Hunter, J.~D. 2007, Computing in Science \& Engineering, 9, 90,
  \dodoi{10.1109/MCSE.2007.55}

\bibitem[{{Kawata} {et~al.}(2019){Kawata}, {Bovy}, {Matsunaga}, \&
  {Baba}}]{kawata_2019}
{Kawata}, D., {Bovy}, J., {Matsunaga}, N., \& {Baba}, J. 2019, \mnras, 482, 40,
  \dodoi{10.1093/mnras/sty2623}

\bibitem[{{Lee} {et~al.}(2019){Lee}, {Beers}, \& {Kim}}]{lee_2019}
{Lee}, Y.~S., {Beers}, T.~C., \& {Kim}, Y.~K. 2019, \apj, 885, 102,
  \dodoi{10.3847/1538-4357/ab4791}

\bibitem[{{Lee} {et~al.}(2017){Lee}, {Beers}, {Kim}, {Placco}, {Yoon},
  {Carollo}, {Masseron}, \& {Jung}}]{lee_2017}
{Lee}, Y.~S., {Beers}, T.~C., {Kim}, Y.~K., {et~al.} 2017, \apj, 836, 91,
  \dodoi{10.3847/1538-4357/836/1/91}

\bibitem[{{Lian} {et~al.}(2020){Lian}, {Thomas}, {Maraston}, {Beers}, {Moni
  Bidin}, {Fern{\'a}ndez-Trincado}, {Garc{\'\i}a-Hern{\'a}ndez}, {Lane},
  {Munoz}, {Nitschelm}, {Roman-Lopes}, \& {Zamora}}]{lian_2020}
{Lian}, J., {Thomas}, D., {Maraston}, C., {et~al.} 2020, \mnras, 497, 2371,
  \dodoi{10.1093/mnras/staa2078}

\bibitem[{{Morrison}(1990)}]{morrison_1990}
{Morrison}, H.~L. 1990, \jrasc, 84, 107

\bibitem[{{Placco} {et~al.}(2014){Placco}, {Frebel}, {Beers}, \&
  {Stancliffe}}]{placco_2014}
{Placco}, V.~M., {Frebel}, A., {Beers}, T.~C., \& {Stancliffe}, R.~J. 2014,
  \apj, 797, 21, \dodoi{10.1088/0004-637X/797/1/21}

\bibitem[{{Sch{\"o}nrich} {et~al.}(2010){Sch{\"o}nrich}, {Binney}, \&
  {Dehnen}}]{schonrich_2010}
{Sch{\"o}nrich}, R., {Binney}, J., \& {Dehnen}, W. 2010, \mnras, 403, 1829,
  \dodoi{10.1111/j.1365-2966.2010.16253.x}

\bibitem[{{Sch{\"o}nrich} {et~al.}(2019){Sch{\"o}nrich}, {McMillan}, \&
  {Eyer}}]{schonrich_2019}
{Sch{\"o}nrich}, R., {McMillan}, P., \& {Eyer}, L. 2019, \mnras, 487, 3568,
  \dodoi{10.1093/mnras/stz1451}

\bibitem[{{Sestito} {et~al.}(2019){Sestito}, {Longeard}, {Martin},
  {Starkenburg}, {Fouesneau}, {Gonz{\'a}lez Hern{\'a}ndez}, {Arentsen},
  {Ibata}, {Aguado}, {Carlberg}, {Jablonka}, {Navarro}, {Tolstoy}, \&
  {Venn}}]{sestito_2019}
{Sestito}, F., {Longeard}, N., {Martin}, N.~F., {et~al.} 2019, \mnras, 484,
  2166, \dodoi{10.1093/mnras/stz043}

\bibitem[{{Sestito} {et~al.}(2020){Sestito}, {Martin}, {Starkenburg},
  {Arentsen}, {Ibata}, {Longeard}, {Kielty}, {Youakim}, {Venn}, {Aguado},
  {Carlberg}, {Gonz{\'a}lez Hern{\'a}ndez}, {Hill}, {Jablonka}, {Kordopatis},
  {Malhan}, {Navarro}, {S{\'a}nchez-Janssen}, {Thomas}, {Tolstoy}, {Wilson},
  {Palicio}, {Bialek}, {Garcia-Dias}, {Lucchesi}, {North}, {Osorio}, {Patrick},
  \& {Peralta de Arriba}}]{sestito_2020}
{Sestito}, F., {Martin}, N.~F., {Starkenburg}, E., {et~al.} 2020, \mnras,
  \dodoi{10.1093/mnrasl/slaa022}

\bibitem[{{Sommer-Larsen} \& {Zhen}(1990)}]{sommer-larsen_1990}
{Sommer-Larsen}, J., \& {Zhen}, C. 1990, \mnras, 242, 10,
  \dodoi{10.1093/mnras/242.1.10}

\bibitem[{{van der Walt} {et~al.}(2011){van der Walt}, {Colbert}, \&
  {Varoquaux}}]{numpy}
{van der Walt}, S., {Colbert}, S.~C., \& {Varoquaux}, G. 2011, Computing in
  Science and Engineering, 13, 22, \dodoi{10.1109/MCSE.2011.37}

\bibitem[{{Virtanen} {et~al.}(2019){Virtanen}, {Gommers}, {Oliphant},
  {Haberland}, {Reddy}, {Cournapeau}, {Burovski}, {Peterson}, {Weckesser},
  {Bright}, {van der Walt}, {Brett}, {Wilson}, {Jarrod Millman}, {Mayorov},
  {Nelson}, {Jones}, {Kern}, {Larson}, {Carey}, {Polat}, {Feng}, {Moore}, {Vand
  erPlas}, {Laxalde}, {Perktold}, {Cimrman}, {Henriksen}, {Quintero}, {Harris},
  {Archibald}, {Ribeiro}, {Pedregosa}, {van Mulbregt}, \&
  {Contributors}}]{scipy}
{Virtanen}, P., {Gommers}, R., {Oliphant}, T.~E., {et~al.} 2019, arXiv
  e-prints, arXiv:1907.10121.
\newblock \doarXiv{1907.10121}

\bibitem[{{Wolf} {et~al.}(2018){Wolf}, {Onken}, {Luvaul}, {Schmidt}, {Bessell},
  {Chang}, {Da Costa}, {Mackey}, {Martin-Jones}, {Murphy}, {Preston}, {Scalzo},
  {Shao}, {Smillie}, {Tisserand}, {White}, \& {Yuan}}]{skymapper_dr1}
{Wolf}, C., {Onken}, C.~A., {Luvaul}, L.~C., {et~al.} 2018, \pasa, 35, e010,
  \dodoi{10.1017/pasa.2018.5}

\bibitem[{{Yanny} {et~al.}(2009){Yanny}, {Rockosi}, {Newberg}, {Knapp},
  {Adelman-McCarthy}, {Alcorn}, {Allam}, {Allende Prieto}, {An}, {Anderson},
  {Anderson}, {Bailer-Jones}, {Bastian}, {Beers}, {Bell}, {Belokurov},
  {Bizyaev}, {Blythe}, {Bochanski}, {Boroski}, {Brinchmann}, {Brinkmann},
  {Brewington}, {Carey}, {Cudworth}, {Evans}, {Evans}, {Gates}, {G{\"a}nsicke},
  {Gillespie}, {Gilmore}, {Nebot Gomez-Moran}, {Grebel}, {Greenwell}, {Gunn},
  {Jordan}, {Jordan}, {Harding}, {Harris}, {Hendry}, {Holder}, {Ivans},
  {Ivezi{\v{c}}}, {Jester}, {Johnson}, {Kent}, {Kleinman}, {Kniazev},
  {Krzesinski}, {Kron}, {Kuropatkin}, {Lebedeva}, {Lee}, {French Leger},
  {L{\'e}pine}, {Levine}, {Lin}, {Long}, {Loomis}, {Lupton}, {Malanushenko},
  {Malanushenko}, {Margon}, {Martinez-Delgado}, {McGehee}, {Monet}, {Morrison},
  {Munn}, {Neilsen}, {Nitta}, {Norris}, {Oravetz}, {Owen}, {Padmanabhan},
  {Pan}, {Peterson}, {Pier}, {Platson}, {Re Fiorentin}, {Richards}, {Rix},
  {Schlegel}, {Schneider}, {Schreiber}, {Schwope}, {Sibley}, {Simmons},
  {Snedden}, {Allyn Smith}, {Stark}, {Stauffer}, {Steinmetz}, {Stoughton},
  {SubbaRao}, {Szalay}, {Szkody}, {Thakar}, {Sivarani}, {Tucker}, {Uomoto},
  {Vanden Berk}, {Vidrih}, {Wadadekar}, {Watters}, {Wilhelm}, {Wyse}, {Yarger},
  \& {Zucker}}]{yanny_2009}
{Yanny}, B., {Rockosi}, C., {Newberg}, H.~J., {et~al.} 2009, \aj, 137, 4377,
  \dodoi{10.1088/0004-6256/137/5/4377}

\bibitem[{{Yoon} {et~al.}(2019){Yoon}, {Beers}, {Tian}, \&
  {Whitten}}]{yoon_2019}
{Yoon}, J., {Beers}, T.~C., {Tian}, D., \& {Whitten}, D.~D. 2019, \apj, 878,
  97, \dodoi{10.3847/1538-4357/ab1ead}

\bibitem[{{Yoon} {et~al.}(2020){Yoon}, {Whitten}, {Beers}, {Lee}, {Masseron},
  \& {Placco}}]{yoon_2020}
{Yoon}, J., {Whitten}, D.~D., {Beers}, T.~C., {et~al.} 2020, \apj, 894, 7,
  \dodoi{10.3847/1538-4357/ab7daf}

\bibitem[{{Yoon} {et~al.}(2016){Yoon}, {Beers}, {Placco}, {Rasmussen},
  {Carollo}, {He}, {Hansen}, {Roederer}, \& {Zeanah}}]{yoon_2016}
{Yoon}, J., {Beers}, T.~C., {Placco}, V.~M., {et~al.} 2016, \apj, 833, 20,
  \dodoi{10.3847/0004-637X/833/1/20}

\bibitem[{{Yoon} {et~al.}(2018){Yoon}, {Beers}, {Dietz}, {Lee}, {Placco}, {Da
  Costa}, {Keller}, {Owen}, \& {Sharma}}]{yoon_2018}
{Yoon}, J., {Beers}, T.~C., {Dietz}, S., {et~al.} 2018, \apj, 861, 146,
  \dodoi{10.3847/1538-4357/aaccea}

\bibitem[{{York} {et~al.}(2000){York}, {Adelman}, {Anderson}, {Anderson},
  {Annis}, {Bahcall}, {Bakken}, {Barkhouser}, {Bastian}, {Berman}, {Boroski},
  {Bracker}, {Briegel}, {Briggs}, {Brinkmann}, {Brunner}, {Burles}, {Carey},
  {Carr}, {Castander}, {Chen}, {Colestock}, {Connolly}, {Crocker}, {Csabai},
  {Czarapata}, {Davis}, {Doi}, {Dombeck}, {Eisenstein}, {Ellman}, {Elms},
  {Evans}, {Fan}, {Federwitz}, {Fiscelli}, {Friedman}, {Frieman}, {Fukugita},
  {Gillespie}, {Gunn}, {Gurbani}, {de Haas}, {Haldeman}, {Harris}, {Hayes},
  {Heckman}, {Hennessy}, {Hindsley}, {Holm}, {Holmgren}, {Huang}, {Hull},
  {Husby}, {Ichikawa}, {Ichikawa}, {Ivezi{\'c}}, {Kent}, {Kim}, {Kinney},
  {Klaene}, {Kleinman}, {Kleinman}, {Knapp}, {Korienek}, {Kron}, {Kunszt},
  {Lamb}, {Lee}, {Leger}, {Limmongkol}, {Lindenmeyer}, {Long}, {Loomis},
  {Loveday}, {Lucinio}, {Lupton}, {MacKinnon}, {Mannery}, {Mantsch}, {Margon},
  {McGehee}, {McKay}, {Meiksin}, {Merelli}, {Monet}, {Munn}, {Narayanan},
  {Nash}, {Neilsen}, {Neswold}, {Newberg}, {Nichol}, {Nicinski}, {Nonino},
  {Okada}, {Okamura}, {Ostriker}, {Owen}, {Pauls}, {Peoples}, {Peterson},
  {Petravick}, {Pier}, {Pope}, {Pordes}, {Prosapio}, {Rechenmacher}, {Quinn},
  {Richards}, {Richmond}, {Rivetta}, {Rockosi}, {Ruthmansdorfer}, {Sand ford},
  {Schlegel}, {Schneider}, {Sekiguchi}, {Sergey}, {Shimasaku}, {Siegmund},
  {Smee}, {Smith}, {Snedden}, {Stone}, {Stoughton}, {Strauss}, {Stubbs},
  {SubbaRao}, {Szalay}, {Szapudi}, {Szokoly}, {Thakar}, {Tremonti}, {Tucker},
  {Uomoto}, {Vanden Berk}, {Vogeley}, {Waddell}, {Wang}, {Watanabe},
  {Weinberg}, {Yanny}, {Yasuda}, \& {SDSS Collaboration}}]{york_2000}
{York}, D.~G., {Adelman}, J., {Anderson}, John~E., J., {et~al.} 2000, \aj, 120,
  1579, \dodoi{10.1086/301513}

\bibitem[{{Yoshii}(1982)}]{yoshii_1982}
{Yoshii}, Y. 1982, \pasj, 34, 365

\bibitem[{{Youakim} {et~al.}(2017){Youakim}, {Starkenburg}, {Aguado}, {Martin},
  {Fouesneau}, {Gonz{\'a}lez Hern{\'a}ndez}, {Allende Prieto}, {Bonifacio},
  {Gentile}, {Kielty}, {C{\^o}t{\'e}}, {Jablonka}, {McConnachie}, {S{\'a}nchez
  Janssen}, {Tolstoy}, \& {Venn}}]{pristine_iii}
{Youakim}, K., {Starkenburg}, E., {Aguado}, D.~S., {et~al.} 2017, \mnras, 472,
  2963, \dodoi{10.1093/mnras/stx2005}

\end{thebibliography}

\end{document}